\documentclass[aps,a4paper,showkeys,nofootinbib,longbibliography,twocolumn]{revtex4-2}
\usepackage{braket}
\usepackage[utf8]{inputenc}
\usepackage{filecontents}
\usepackage{natbib}
\usepackage{amsmath,amssymb,bm,amsthm}
\usepackage{subfigure}
\usepackage{graphicx}
\usepackage{dcolumn}
\usepackage{bm}
\usepackage[mathlines]{lineno}
\usepackage{booktabs}
\usepackage{color}
\newcounter{one}
\usepackage{url}

\usepackage{textcase}

\usepackage{colortbl}
\usepackage{tabularx}
\usepackage{verbatim}
\usepackage{multirow}

\usepackage[T1]{fontenc} 
\usepackage{lmodern}
\usepackage{bbm}
\usepackage[utf8]{inputenc}
\usepackage{amsfonts}
\usepackage{array}

\usepackage{bbm}
\usepackage{enumitem}
\usepackage{umoline}
\usepackage[usenames,svgnames]{xcolor}
\usepackage{natbib}
\usepackage[hyperindex,breaklinks]{hyperref}
\hypersetup{
     colorlinks=true,       		
     linkcolor=Navy,          	
     citecolor=Navy,            
     filecolor=Navy,      		
     urlcolor=Navy,           	
    runcolor=cyan,
 }
\usepackage{graphicx}
\usepackage{amsfonts}
\usepackage{amssymb}
\usepackage{amsmath}
\usepackage{bbm}
\usepackage{enumitem}

\newcommand{\tr}[0]{ {\rm tr}}

\newcommand{\half}[1]{{ \rm h}}
\newcommand{\Oorderof}{\mathcal{O}}
\newcommand{\orderof}[1]{\Oorderof(#1)} 

\newcommand{\for}[0]{\quad \textrm{for} \quad}

\newcommand{\poly}{{\rm poly}}

\def\beq{\begin{equation}}
\def\eeq{\end{equation}}
\def\nbeq{\begin{equation*}}
\def\neeq{\end{equation*}}
\def\<{\langle}
\def\>{\rangle}

\def\tr{{\rm tr}}

\newcommand{\mA}{{\mathcal{A}}}
\newcommand{\mB}{{\mathcal{B}}}

\newtheorem{theorem}{Theorem}

\newtheorem{lemma}{Lemma}

\newtheorem{prop}[theorem]{Proposition} 

\newcommand{\br}[1]{\left( #1 \right)}
\newcommand{\brr}[1]{\left[ #1 \right]}

 \newcommand{\norm}[1]{\left \|  #1 \right \|}

\newcommand{\abs}[1]{\left| #1 \right|}

\usepackage{mathtools}
\def\multiset#1#2{\ensuremath{\left(\kern-.3em\left(\genfrac{}{}{0pt}{}{#1}{#2}\right)\kern-.3em\right)}}

\renewcommand\thefootnote{*\arabic{footnote}}

\begin{document}



\title{Provably Efficient Simulation of 1D Long-Range Interacting Systems \\ at Any Temperature}

\author{Rakesh Achutha$^{1,2}$}
\email{achutha.rakesh.mat20@itbhu.ac.in}

\author{Donghoon Kim$^1$}
\email{donghoon.kim@riken.jp}

\author{Yusuke Kimura$^1$}
\email{yusuke.kimura.cs@riken.jp}

\author{Tomotaka Kuwahara$^{1,3,4}$}
\email{tomotaka.kuwahara@riken.jp}

\affiliation{$^{1}$
Analytical quantum complexity RIKEN Hakubi Research Team, RIKEN Center for Quantum Computing (RQC), Wako, Saitama 351-0198, Japan
}
\affiliation{$^{2}$
Department of Computer Science and Engineering, Indian Institute of Technology (Banaras Hindu University), Varanasi, 221005, India
}

\affiliation{$^{3}$
PRESTO, Japan Science and Technology (JST), Kawaguchi, Saitama 332-0012, Japan}

\affiliation{$^{4}$
RIKEN Cluster for Pioneering Research (CPR), Wako, Saitama 351-0198, Japan
}

\begin{abstract}
We introduce a method that ensures efficient computation of one-dimensional quantum systems with long-range interactions across all temperatures. Our algorithm operates within a quasi-polynomial runtime for inverse temperatures up to $\beta=\poly(\ln(n))$. At the core of our approach is the Density Matrix Renormalization Group algorithm, which typically does not guarantee efficiency. We have created a new truncation scheme for the matrix product operator of the quantum Gibbs states, which allows us to control the error analytically. Additionally, our method can be applied to simulate the time evolution of systems with long-range interactions, achieving significantly better precision than that offered by the Lieb-Robinson bound.
\end{abstract}

\maketitle

\textit{Introduction.---} 
Unraveling the complex patterns of quantum many-body systems presents one of the most significant challenges in modern physics.
Among these challenges, characterizing the thermal equilibrium quantum state (also known as the quantum Gibbs state) at zero and non-zero temperatures stands as a central target. These states are generally difficult to manage in high-dimensional systems, as mentioned in previous studies~\cite{BULATOV2005148,Goldberg13161}. However, a variety of techniques have been developed to address their properties in one-dimensional systems~\cite{PhysRevB.76.201102,PhysRevB.91.045138,kim2017markovian,kuwahara2018polynomialtime,PhysRevX.11.011047,PRXQuantum.2.040331,Fawzi2023subpolynomialtime,gondolf2024CMI,Scalet_2024}. 
Notably, techniques such as the transfer matrix method used in classical Ising models~\cite{RevModPhys.39.883} and its generalization to 1D quantum systems~\cite{Araki1969} have been well-recognized. More recently, the Density Matrix Renormalization Group (DMRG) algorithm has emerged as a promising approach~\cite{Foot0}, enabling the numerical construction of the matrix product operator (MPO) that encapsulates all information about the equilibrium states~\cite{PhysRevLett.69.2863,PhysRevB.72.220401,PhysRevLett.93.207204,PhysRevA.78.022103,PhysRevLett.102.190601,Stoudenmire_2010,PhysRevD.92.034519,PhysRevB.94.235142,PhysRevD.94.085018,Chen2018,PhysRevB.100.045110,PAECKEL2019167998,PhysRevX.10.031040,PhysRevLett.125.170604,RevModPhys.77.259,PhysRevB.103.205128,PhysRevB.106.165126,10.21468/SciPostPhys.14.4.085}.  
Despite its promise, providing a rigorous justification for the precision guarantee in the DMRG (or MPO-based) algorithm remains a critical open problem at both zero and non-zero temperatures;  
for example, the exponential tensor renormalization group (XTRG) algorithm~\cite{Chen2018} uses a variational approach to construct the Gibbs state but lacks precision guarantees.
Nevertheless, several methods have been proposed to develop a `DMRG-type' algorithm to construct the MPO with guaranteed accuracy and time efficiency~\cite{Arad2017,PhysRevX.11.011047}.

In one-dimensional systems at non-zero temperatures, the construction of the MPO for short-range interacting systems has received extensive attention. The pioneering method, based on the cluster expansion technique by Molnar et al.~\cite{PhysRevB.91.045138}, introduced a polynomial-time algorithm. This algorithm works with a runtime scaling as $(n/\epsilon)^{\mathcal{O}(\beta)}$, where $n$ is the system size, to approximate the 1D quantum Gibbs state by the MPO within an error $\epsilon$. Subsequent advancements have refined the error bounds, with the leading-edge approach achieving a runtime of $e^{\tilde{\mathcal{O}}(\beta) + \tilde{\mathcal{O}}(\sqrt{\beta \ln(n/\epsilon)})}$~\cite{PhysRevX.11.011047,PRXQuantum.2.040331}. This technique utilizes the imaginary-time adaptation of the Haah-Hastings-Kothari-Low (HHKL) decomposition, originally devised for real-time evolution~\cite{HHKL}.

Extending from the study of short-range interactions in one-dimensional (1D) systems, we explore the realm of long-range interactions, characterized by power-law decay. These interactions impart high-dimensional traits to systems, manifesting behaviors not observed in short-range interacting systems. Notably, phenomena such as phase transitions, typically associated with higher dimensions, can occur even in 1D systems with long-range interactions~\cite{Dyson1969,PhysRevLett.37.1577,PhysRev.187.732,PhysRevB.64.184106,Maity_2020,PhysRevB.107.L140410,schuckert2023}. 
Additionally, long-range interactions are increasingly relevant in contemporary many-body physics experiments, such as those involving ultracold atomic systems~\cite{RevModPhys.80.885,RevModPhys.82.2313,Islam583,bernien2017probing,zhang2017observation,Neyenhuise1700672}, highlighting their practical significance~\cite{RevModPhys.95.035002}.

Beyond the realm of short-range interacting systems, the complexity analysis of long-range interactions presents substantial challenges, especially at low temperatures. Most existing methods are tailored specifically for short-range interactions, and their direct application to long-range systems often results in a loss of efficiency~\cite{Foot5}. While the cluster expansion method proves effective at high temperatures~\cite{mann2020efficient,PRXQuantum.5.010305,PRXQuantum.4.020340}, providing an efficient algorithm to analyze the properties of the quantum Gibbs state in systems with long-range interactions, it falls short at low temperatures. Notably, it does not offer an approximation by the MPO, leaving the computational approach at low temperatures as an unresolved issue.

In this letter, we introduce an efficiency-guaranteed method for constructing the MPO based on the DMRG algorithm. Our method achieves a runtime of:
\begin{align}
&e^{\mathcal{O}(\beta\ln^3(n/\epsilon))} \quad \text{(general cases)},
\end{align}
and for Hamiltonians that are $2$-local, as defined in Eq.~\eqref{2-local_cases} below, the runtime improves to:
\begin{align}
e^{\tilde{\mathcal{O}}(\beta\ln^2(n/\epsilon))} \quad \text{(2-local cases)}.
\end{align}
The results hold for arbitrary inverse temperature $\beta$. 
In particular, as long as $\beta=\poly(\ln(n))$ [i.e., for values of $\beta$ up to $\poly(\ln(n))$], these results demonstrate a quasi-polynomial time complexity. 
Notably, in this temperature regime, the quantum Gibbs states typically approximate the non-critical ground states~\cite{PhysRevB.76.035114,PhysRevLett.124.200604}.
Furthermore, while our primary focus has been on constructing MPOs for quantum Gibbs states, our approach is equally applicable to real-time evolutions. 
To our knowledge, this is the first rigorous justification of a classical simulation method with guaranteed efficiency for managing the real and imaginary-time dynamics of long-range interacting Hamiltonians.

\begin{figure*}[tt]
\centering
\includegraphics[clip, scale=0.32]{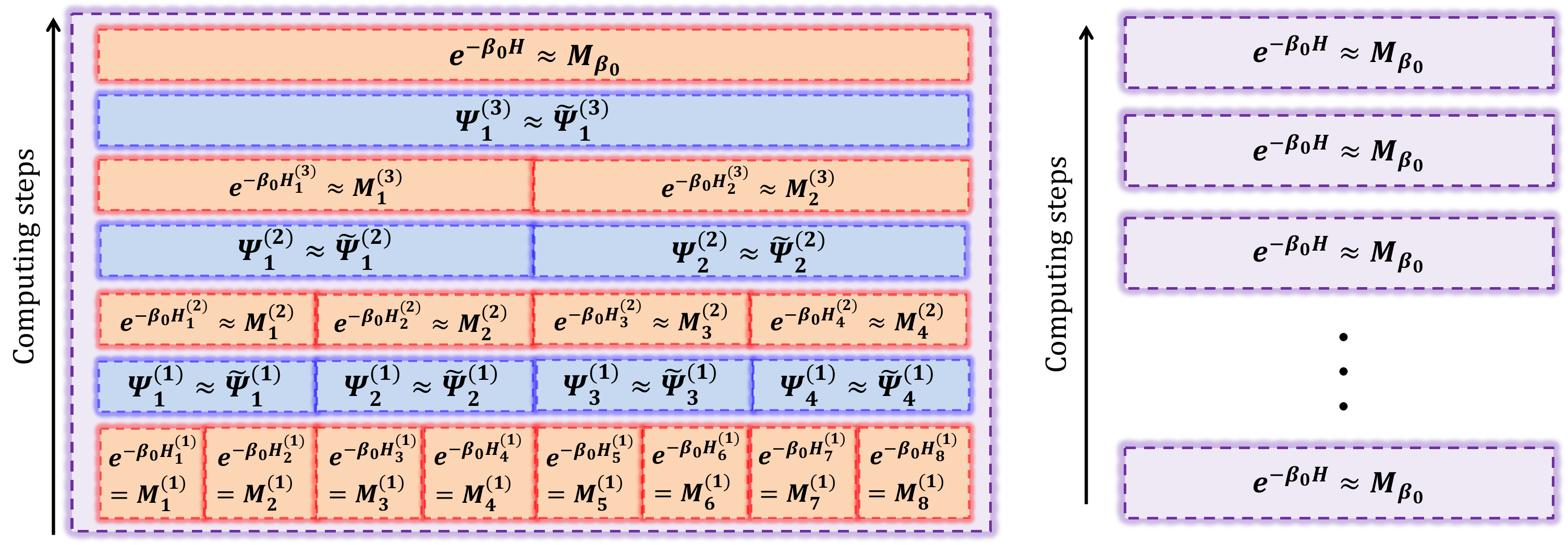}
\caption{Illustration of the Matrix Product Operator (MPO) construction for approximating the quantum Gibbs state. (a) \textit{High-Temperature Quantum Gibbs States:} The process begins with the product state of quantum Gibbs states across eight blocks. In the initial merging process $\{\Psi_1^{(1)},\Psi_2^{(1)},\Psi_3^{(1)},\Psi_4^{(1)}\}$, the block size is doubled, yielding quantum Gibbs states $\{e^{-\beta_0 H_1^{(2)}},e^{-\beta_0 H_2^{(2)}},e^{-\beta_0 H_3^{(2)}},e^{-\beta_0 H_4^{(2)}}\}$. After $q$ merging processes, with each process doubling the block size, the final block size becomes $2^q$ times the original. To cover the total system size $n$, $\log_{2}(n)$ merging processes are required. Each merging operator $\{\Psi_{s}^{(q)}\}_{s,q}$ is approximated by MPOs $\{\tilde{\Psi}_{s}^{(q)}\}_{s,q}$ with truncated bond dimensions [see Eq.~\eqref{tilde_Psi_M_0}], facilitating the construction of the MPO $M_{\beta_0}$ for the high-temperature quantum Gibbs state $e^{-\beta_0 H}$. Approximation errors are iteratively estimated as detailed in Proposition~2, noting that all approximation errors stem from these merging operators. (b) \textit{Low-Temperature Quantum Gibbs States:} By iteratively combining high-temperature quantum Gibbs states, the Gibbs state at any desired temperature $\beta$ is achieved by exponentiating $e^{-\beta_0 H}$ to $\beta/\beta_0$, where $\beta_0$ is chosen so that $\beta/\beta_0$ is an integer. The approximation error between $e^{-\beta H}$ and $(M_{\beta_0})^{\beta/\beta_0}$ is managed by controlling the error between $e^{-\beta_0 H}$ and $M_{\beta_0}$ in terms of an arbitrary Schatten $p$-norm, as achieved in~\eqref{Schatten_p/high/temp}.}
\label{fig:Fig1}
\end{figure*}

In terms of technical advancements, the traditional DMRG algorithm faced challenges in accurately estimating the truncation errors of bond dimensions during iterative steps~\cite{PhysRevB.72.220401,PhysRevLett.93.207204} (see also Ref.~\cite[Sec. IV C therein]{PhysRevX.11.011047}). To address this, we developed a new algorithm that captures the core mechanics, as illustrated in Fig.~\ref{fig:Fig1}. Our approach comprises two main steps: (i) constructing the MPO at high temperatures with a precisely controlled error, using arbitrary Schatten $p$-norms, and (ii) merging high-temperature quantum Gibbs states into their low-temperature counterparts. The success of the first step allows us to apply techniques from Refs.~\cite{PhysRevB.76.201102,PhysRevX.11.011047} in the second step.
Hence, the primary challenge lies in the construction of the high-temperature MPO in the first step. We begin by constructing the quantum Gibbs state in independent small blocks and iteratively merge them into larger ones. 
This approach retains the spirit of the original DMRG algorithm developed by White~\cite{PhysRevLett.69.2863}, focusing on local interactions and iterative improvements.
A key innovation of our algorithm is the efficient control of the precision in the merging process, quantified by the general Schatten $p$-norm (see Proposition~2 below).

\textit{Setup.---} We consider a one-dimensional quantum system composed of $n$ qudits, each in a $d$-dimensional Hilbert space. The total set of sites in this system is denoted by $\Lambda$, i.e., $\Lambda = \{1, 2, 3, \dots, n\}$. We define a $k$-local Hamiltonian $H$ for this system by:
\begin{align}
    H = \sum_{|Z| \leq k} h_Z, \quad \max_{i \in \Lambda} \sum_{Z : Z \ni i} \|h_Z\| \leq g, \label{eq:Hdef}
\end{align}
where $Z \subset \Lambda$, and $|Z|$ represents the cardinality of $Z$, meaning the number of sites involved in each interaction term $h_Z$. The term $\|\cdots\|$ denotes the operator norm, which is the maximum singular value of the operator. For any arbitrary subset $L \subseteq \Lambda$, the subset Hamiltonian $H_L$ is defined to include only those interaction terms involving sites in $L$, specifically:
\begin{align}
    H_L = \sum_{Z:Z \subset L} h_Z. \notag 
\end{align}

To elucidate the assumptions underpinning our analysis, let us consider a decomposition of the total system $\Lambda$ into subsets $A$ and $B$ such that $\Lambda = A \sqcup B$. Here, subset $A$ is defined as $[i, i']\cap \Lambda$ with $1 \leq i < i' \leq n$. We assume that the norm of the boundary interaction between $A$ and $B$, denoted by $\|\partial h_A\|$, is finite:
\begin{align}
    \|\partial h_A\| \leq \tilde{g}, \quad \forall A, \quad \partial h_A := H - (H_A + H_B), \label{basic_assumption_main}
\end{align}
where $\tilde{g}$ represents an $\mathcal{O}(1)$ constant. This condition is more general than the assumption of power-law decay of interactions. If we specifically consider the power-law decay condition in the form:
\begin{align}
    \sum_{Z: Z \ni \{i, i'\}} \|h_Z\| \leq \frac{J}{|i - i'|^\alpha} \for i\neq i',
\end{align}
then the condition in \eqref{basic_assumption_main} is satisfied as long as $\alpha > 2$
(see Supplementary materials~\cite{Supplement_thermal}).

Throughout the paper, we analyze the quantum Gibbs state as follows:
\begin{align}
    \rho_{\beta} := \frac{1}{Z_\beta}e^{-\beta H}, \quad Z_\beta := \operatorname{tr}\left(e^{-\beta H}\right). \notag 
\end{align}
We aim to approximate the Gibbs state \(\rho_\beta\) by a MPO, which is generally described as follows:
\begin{align}
    M = \sum_{\substack{s_1, s_2, \dots, s_n = 1 \\ s_1', s_2', \dots, s_n' = 1}}^d &\operatorname{tr}\left(M_1^{[s_1, s_1']} M_2^{[s_2, s_2']} \cdots M_n^{[s_n, s_n']}\right) \notag \\
    &\left|s_1, s_2, \dots, s_n\right\rangle \left\langle s_1', s_2', \dots, s_n'\right|,
\end{align}
where each $\{M_j^{[s_j, s_j']}\}_{j,s_{j},s_{j}'}$ is a $D \times D$ matrix, with $D$ referred to as the bond dimension.
As measures of the approximation, we often use the Schatten $p$-norm as follows:
$
\norm{O}_p :=[\tr(|O|^p)]^{1/p}
$
with $O$ an arbitrary operator and $|O| = \sqrt{O^{\dagger}O}$. For $p=1$, it is equivalent to the trace norm, and for $p=\infty$, it corresponds to the operator norm.

We denote the bond dimension of the MPO form of the Hamiltonian by $D_H$, which is at most of $n^k d^k$ in general~\cite{Foot1}. 
In particular, if we restrict ourselves to a 2-local Hamiltonian in the form of:
\begin{align}
    H &= \sum_{i < i'} \frac{1}{|i - i'|^\alpha} \sum_{\xi, \xi' = 1}^{d^2} J_{\xi,\xi'} P_{i,\xi} \otimes P_{i',\xi'}, \label{2-local_cases}
\end{align}
where $\{P_{i,\xi}\}_{\xi=1}^{d^2}$ denotes the operator bases at site $i$ and $J_{\xi,\xi'}$ are the corresponding coefficients, the Hamiltonian can be approximated by an MPO with bond dimension $D_H = c_H \ln^2(n/\epsilon)$, where $c_H = \mathcal{O}(1)$, up to an error of $\epsilon$ (see Supplementary material~\cite[Lemma~2 therein]{Supplement_thermal}).
We note that any subset Hamiltonian $H_L$ is also described by an MPO with the bond dimension $D_H$ of the global Hamiltonian.

\textit{Main Result.---} 
Our task is to develop a time-efficient method using an MPO to approximate the quantum Gibbs state for a given $\beta$ within a desired error $\epsilon$.
The main result concerning the time-efficient construction of the MPO is as follows:

{~}

\noindent
\textbf{Theorem.} 
For an arbitrary $\beta$, we can efficiently compute the MPO $M_\beta$ that approximates the unnormalized Gibbs state up to an error $\epsilon$:
\begin{align}
    \label{Main_inequ_for/MPO_ap}
    \norm{e^{-\beta H} - M_\beta}_p \leq \epsilon \norm{e^{-\beta H}}_p, \quad \forall p.
\end{align}
The bond dimension of $M_\beta$ and the time complexity are given by 
$e^{C_0 \beta \ln^3(n/\epsilon)}$,
where $C_0$ is an $\mathcal{O}(1)$ constant. 
Typically, we consider $\epsilon = 1/\poly(n)$, which leads to a time complexity of $e^{\mathcal{O}(\beta) \ln^3(n)}$. 
If we consider the 2-local Hamiltonian in Eq.~\eqref{2-local_cases}, 
the time complexity qualitatively improves to $e^{C_0 \beta\ln^2(n/\epsilon) \ln\ln (n/\epsilon)}$.
From the theorem, we can efficiently simulate the quantum Gibbs state in long-range interacting systems~\cite{Foot3}.
Finally, we emphasize that the preparation of the quantum Gibbs state on a quantum computer is also possible by a quasi-polynomial time complexity~\cite{Supplement_thermal}.

{~}

In the following, we detail the explicit calculation steps for constructing the MPO $M_\beta$, which is also illustrated in Fig.~\ref{fig:Fig1}:
\begin{enumerate}
    \item \textit{Initial State Preparation:} We begin with a product state of the quantum Gibbs states of small blocks, each containing $2$ sites, at an inverse temperature $\beta_0$. This temperature $\beta_0$ is chosen to be smaller than $1/(24gk^2)$, which will be utilized to ensure the convergence of the expansion~\eqref{tilde_Psi_M_0}. The exact MPO representation of the initial state has a bond dimension at most $d^2$, where $d$ is the dimension of the local Hilbert space.
    
    \item \textit{Merging Operation:} Adjacent quantum Gibbs states on blocks are merged using a merging operator $\Psi$. This operator is approximated by $\tilde{\Psi}$ using truncated expansions as described in Eq.~\eqref{tilde_Psi_M_0}, up to $m_0$-th order, where $m_0 = \ln(n/\epsilon)$. Consequently, the bond dimension of the MPO form of $\tilde{\Psi}$ becomes $D_H^{\orderof{m_0}}$, with $D_H$ representing the bond dimension of the Hamiltonian.
    
    \item \textit{Approximation of High-Temperature Gibbs State:} By repeating the merging processes $\log_2(n)$ times, we achieve the global quantum Gibbs state $e^{-\beta_0 H}$. This high-temperature quantum Gibbs state is approximated by the MPO $M_{\beta_0}$, which has a bond dimension $D_H^{\orderof{m_0 \ln(n)}}$.

    \item \textit{Approximation of Low-Temperature Gibbs State:} The high-temperature quantum Gibbs state is connected $(\beta/\beta_0)$ times to yield the MPO $(M_{\beta_0})^{\beta/\beta_0}$, serving as the approximation of the desired quantum Gibbs state $e^{-\beta H}$. In this process, we need to utilize the error bound in terms of Schatten $p$-norm (see Lemma~3 below).
\end{enumerate}
\noindent
In the next section, we will focus on analytically estimating the precision error and the time complexity resulting from these approximations.

\textit{Proof of main theorem.---} 
We here provide the outline of the proof, and we defer the details to Supplementary materials~\cite{Supplement_thermal}.
We follow Steps 1 to 4 above and estimate the precision error depending on the bond dimension. 
We denote the Gibbs state of the block $L_s^{(q)}$ as $e^{-\beta_{0} H_s^{(q)}}$, where $H_s^{(q)}$ serves as an abbreviation for the subset Hamiltonian $H_{L_s^{(q)}} = \sum_{Z : Z \subset L_s^{(q)}} h_Z$ on block $L_s^{(q)}$. Under this notation, $e^{-\beta_{0} H_1^{(q_0)}} = e^{-\beta_{0} H}$ represents the Gibbs state of the entire system, encompassing all interactions within the system.

We then estimate the error arising from the merging operations. 
In general, we define a merging operator $\Psi$ that connects two subsets, $A$ and $B$, as follows:
\begin{align}
\label{merging/Psi/op}
    \Psi = e^{-\beta_{0} H_{AB}} e^{\beta_{0} (H_A + H_B)},
\end{align}
which implies the relation $e^{-\beta_{0} H_{AB}} = \Psi e^{-\beta_{0} H_A} e^{-\beta_{0} H_B}$. Here, $\Psi$ facilitates the merging of the Gibbs states of $A$ and $B$ to form the combined Gibbs state on the larger block $AB$, as illustrated in Fig.~\ref{fig:Fig1}(a). 
In this context, $H_{AB} = H_A + H_B + \partial h_A$, where $\partial h_A$ represents the boundary interactions between systems $A$ and $B$, as defined in Eq.~\eqref{basic_assumption_main}. 

To construct an MPO with a small bond dimension for $\Psi$, we first approximate $\Psi$ using a polynomial expansion in terms of $\beta_{0}$. 
However, the independent approximations of $e^{-\beta_{0} H_{AB}}$ and $e^{\beta_{0} (H_A + H_B)}$ require polynomial degrees of at least $(\beta_{0} \norm{H_{AB}})^{1/2}$ and $(\beta_{0} \norm{H_A} + \beta_{0} \norm{H_B})^{1/2}$, respectively~\cite{TCS-065}. This results in a demand for sub-exponentially large bond dimensions for an accurate approximation.

The key idea here is that by combining $e^{-\beta_{0} H_{AB}}$ and $e^{\beta_{0} (H_A + H_B)}$ the truncation order is significantly reduced for a good approximation of $\Psi$. 
We adopt the following polynomial approximation of the merging operator~\eqref{merging/Psi/op}: 
\begin{align}
\label{tilde_Psi_M_0}
    \Tilde{\Psi}  = \sum^{m_0}_{m=0} \beta_{0}^{m} \sum_{s_1+s_2=m} (-1)^{s_1} \frac{H_{AB}^{s_1}(H_A+H_B)^{s_2}}{s_1!s_2!},
\end{align}
where we apply the Taylor expansion for each of $e^{-\beta_{0} H_{AB}}$ and $e^{\beta_{0} (H_A + H_B)}$ and truncate the terms with $\beta_{0}^m$ for $m>m_0$. 
Then, we prove the following precision error:  \\
\noindent
\textbf{Proposition 1.}  If $\beta_0$ is smaller than $1/(24gk^2)$, we achieve the approximation of 
\begin{align}
\|\Psi - \Tilde{\Psi}\| \leq \delta_0 \for m_0\ge \log_2 \br{c_0/\delta_0} , 
\end{align}
with $c_0=e^{\tilde{g}/(6gk^2)}$ which is an $\orderof{1}$ constant.

To derive the MPO form of \(\Tilde{\Psi}\), we first create individual MPOs for each term in the expansion \eqref{tilde_Psi_M_0}. These individual MPOs are then combined to obtain the MPO representation of \(\Tilde{\Psi}\).
Given the MPO form of the Hamiltonians \(H_{AB}\) and \(H_A + H_B\), the bond dimension required for the term \(H_{AB}^{s_1}(H_A + H_B)^{s_2}\) is at most \(D_H^{s_1}D_H^{s_2} = D_H^m\), where \(s_1 + s_2 = m\). The total bond dimension required to represent all terms up to \(m_0\) in the expansion is calculated as:
\begin{align}
\label{define_bond_tilde_psi}
\sum_{m=0}^{m_0} (m+1) D_H^m \leq (m_0+1)^2 D_H^{m_0} =: \tilde{D}_{\delta_0}.
\end{align}
Thus, we construct the MPO with the bond dimension \(\tilde{D}_{\delta_0}\) to approximate the merging operator \(\Psi\) for any arbitrary adjacent subsets \(A\) and \(B\).

After constructing the MPO approximation of the merging operator, we proceed to find the MPO for the combined Gibbs state. 
We in general consider the merging of $e^{-\beta_{0} H_{2s-1}^{(q-1)}}$ and $e^{- \beta_{0} H_{2s}^{(q-1)}}$, which is given by 
$e^{-\beta_{0} H_{s}^{(q)}}= \Psi_{s}^{(q-1)} e^{-\beta_{0} H_{2s-1}^{(q-1)}} e^{-\beta_{0} H_{2s}^{(q-1)}}$. 
Our purpose is to estimate the approximation by $\tilde{\Psi}_{s}^{(q-1)} M_{2s-1}^{(q-1)} M_{2s}^{(q-1)}$ when approximate MPOs $M_{2s-1}^{(q-1)}$ and $M_{2s}^{(q-1)}$ are given for $e^{-\beta_{0}H_{2s-1}^{(q-1)}}$ and $e^{-\beta_{0} H_{2s}^{(q-1)}}$, respectively. 
For simplicity, we denote $\Psi_{s}^{(q-1)}$ as $\Psi$ and $\tilde{\Psi}_{s}^{(q-1)}$  as $\tilde{\Psi}$, by omitting the indices $s$ and $q$.  
 \\
\noindent
\textbf{Proposition 2.} Let $M_{2s - 1}^{(q - 1)}$ and $M_{2s}^{(q - 1)}$ represent the MPOs approximating $e^{-\beta_{0} H_{2s - 1}^{(q-1)}}$ and $e^{-\beta_{0} H_{2s}^{(q-1)}}$, respectively, such that
$$
\norm{e^{-\beta_{0} H_{i}^{(q-1)}} - M_{i}^{(q-1)}}_p \leq \epsilon_{q-1} \norm{e^{-\beta_{0} H_{i}^{(q-1)}}}_p 
$$
$\forall p$ with $i = 2s-1,2s$.
Using the approximate merging operator $\Tilde{\Psi}$ with $\|\Psi - \Tilde{\Psi}\| \leq \delta_0$, the merged quantum Gibbs state $e^{-\beta_{0} H_s^{(q)}}$ can be approximated as:
\begin{align}
\norm{e^{-\beta_{0} H_{s}^{(q)}} - M_{s}^{(q)}}_p \leq &\epsilon_{q} \norm{e^{-\beta_{0} H_{s}^{(q)}}}_p, \notag 
\end{align}
with $\epsilon_q = a_2 \delta_0 + a_1 \epsilon_{q-1}$,  
where the MPO $M_s^{(q)}$ is constructed as $M_s^{(q)} = \Tilde{\Psi} M_{2s-1}^{(q-1)} M_{2s}^{(q-1)}$, and $a_1$ and $a_2$ are $\mathcal{O}(1)$ constants with $a_1 > 1$.

This shows that $\epsilon_q$ and $\epsilon_{q-1}$ are related by a linear equation, leading to a recursive relation on $q$. By solving this recursive relation, we find:
$
    \epsilon_q  \leq a_2 \delta_0 q a_1^{q-2},
$
using the fact that $\epsilon_1 = 0$, because the MPO description of the initial state is exact.
We then obtain the MPO approximation $M_{\beta_0}$ of the Gibbs state $e^{-\beta_0 H}$, which results in the following error $\forall p$:
\begin{align}
    \norm{e^{-\beta_0 H} - M_{\beta_0}}_p \leq \epsilon'_0 \norm{e^{-\beta_0 H}}_p
    \label{Schatten_p/high/temp}
\end{align}
with $\epsilon'_0:= a_2 \delta_0 q_0 a_1^{q_0-2}$.
The MPO $M_{\beta_0} = M_1^{(q_0)}$ is constructed through $q_0$ approximate merging processes $\Tilde{\Psi}$, starting from the Gibbs states in layer 1. Consequently, the bond dimension increases by a factor of $\tilde{D}_{\delta_0}^{q_0}$, where $\tilde{D}_{\delta_0}$ was defined as the bond dimension of $\Tilde{\Psi}$ in Eq.~\eqref{define_bond_tilde_psi}.
If we consider the error $\|\Psi - \Tilde{\Psi}\| \leq \delta_0 = \epsilon/\mathrm{poly}(n)$, then the bond dimension becomes $\tilde{D}_{\delta_0}^{q_0} = D_H^{\orderof{\ln^2 (n/\epsilon)}}$, as $q_0 = \log_2(n)$ and $\tilde{D}_{\delta_0} = D_H^{\orderof{\ln (1/\delta_0)}}$.

Finally, we use the MPO $M_{\beta_0}$ for the high-temperature Gibbs state $e^{-\beta_0 H}$ to approximate the target low-temperature Gibbs state $e^{-\beta H}$, as illustrated in Fig.~\ref{fig:Fig1}(b). We combine the $(\beta/\beta_0)$ MPOs $M_{\beta_0}$ to approximate the low-temperature Gibbs state, resulting in:
$
(M_{\beta_0})^{\beta/\beta_0} \approx e^{-\beta H}.
$
Consequently, the bond dimension is multiplied $(\beta/\beta_0)$ times, leading to a final required bond dimension of the order $D_H^{\orderof{\beta \ln^2 (n/\epsilon)}}$.

To derive the error bound, we use the statement from \cite[ineq. (C6) therein]{PhysRevX.11.011047}:\\
{\bf Lemma~3.}  
Let $p_1$ and $p_2$ be arbitrary positive integers.
Then, if 
$
\|e^{-\beta_0H} - M_{\beta_0}\|_{p_1p_2} \leq \epsilon' \|e^{-\beta_0H}\|_{p_1p_2},
$
The inequality of 
\begin{align}
\norm{e^{-p_1\beta_0H} - M_{\beta_0}^{p_1}}_{p_2} \leq (3e/2)p_1\epsilon' \norm{e^{-p_1\beta_0H}}_{p_2}
\end{align}
holds for arbitrary $\beta_0$ and $H$. 

We have already derived the MPO approximation for a general Schatten $p$-norm in~\eqref{Schatten_p/high/temp}. 
Therefore, the above lemma yields 
\begin{align}
    \norm{e^{-\beta H} - (M_{\beta_0})^{\beta/\beta_0}}_p \leq \epsilon_{\beta} \norm{e^{-\beta H}}_p,
\end{align}
where 
$
\epsilon_{\beta} = 5(\beta/\beta_0)  \epsilon'_0.
$
Now, let us choose $\beta_0=\beta/\lceil 24gk^2\beta \rceil$. Then, because of $a_1^{q_0} = \mathrm{poly}(n)$ with $q_0 = \log_2(n)$, the choice of $\delta_0 = \epsilon/\mathrm{poly}(n)$ yields the main precision bound in Eq.~\eqref{Main_inequ_for/MPO_ap}. Here, we let $M_\beta = (M_{\beta_0})^{\beta/\beta_0}$ and assume $\beta<n$ without loss of generality~\cite{Foot2}.

By estimating each of the algorithm steps, we can see that the runtime is also upper-bounded by $D_H^{\orderof{\beta \ln^2 (n/\epsilon)}}$ (see Supplementary materials~\cite{Supplement_thermal}).
We thus prove the main theorem.  $\square$

\textit{Extension to Real-Time Evolution--}
In our method, we did not rely on the fact that the inverse temperature $\beta$ is a real number. Therefore, it is possible to generalize our approach to cases where $\beta$ is a complex number. By considering the case where $\beta = it$, we can address the simulation of real-time evolution. 

Given an MPO $M_t$ that approximates $e^{-iHt}$, we can estimate the error as:
$\norm{e^{-iHt}\ket{\Psi} - M_t\ket{\Psi}} \leq \norm{e^{-iHt} - M_t}_\infty$ for an arbitrary quantum state $\ket{\Psi}$. 
By following our algorithm, we can construct an MPO $M_t$ for $e^{-iHt}$ that satisfies the error bound $\norm{e^{-iHt} - M_t}_\infty \leq  \epsilon \norm{e^{-iHt}}_\infty=\epsilon$. The bond dimension and time complexity required are given by a quasi-polynomial form of $e^{\orderof{t\ln^3 (n/\epsilon)}}$. 
Compared to methods based on the Lieb-Robinson bound, our approach significantly reduces the time complexity. Assuming the Lieb-Robinson bound as $\norm{[O_i(t), O_j]} \leq t^{\eta}/r^\alpha$ for $\alpha > 2$~\cite{PhysRevLett.114.157201,PhysRevX.10.031010,kuwahara2020absence,PhysRevLett.127.160401}, the time cost to calculate the average value of a time-evolved local observable $O_i(t)$ scales as $e^{(t^{\eta}/\epsilon)^{1/(\alpha-1)}}$.

\textit{Summary and outlook.---}
In this letter, we have presented an algorithm that achieves quasi-polynomial time complexity in approximating the quantum Gibbs state of 1D long-range interacting systems using MPOs. This was accomplished by adopting a DMRG-type method, as depicted in Fig.~\ref{fig:Fig1}. 
We note that achieving quasi-polynomial complexity is likely the best attainable result in general~\cite{Foot4}. 
However, there is hope to achieve polynomial time complexity for the $2$-local cases, as described in Eq.~\eqref{2-local_cases}. Identifying the optimal time complexity for simulating long-range interacting systems remains an important open question.

Another intriguing avenue for exploration is whether our method can be extended to cases where the power-law decay is slower than $r^{-2}$ ($\alpha < 2$), where the assumption of finite boundary interactions in Eq.~\eqref{basic_assumption_main} may no longer hold. In specific cases, this condition can still be recovered. For instance, in fermion systems with long-range hopping and short-range fermion-fermion interactions, the condition in Eq.~\eqref{basic_assumption_main} holds for $\alpha > 3/2$~\cite{Kuwahara2020arealaw}. Another interesting class is the non-critical quantum Gibbs state, where power-law clustering is satisfied. In such systems, the MPO approximation is expected to hold for $\alpha > 1$, since the mutual information for any bipartition obeys the area law~\cite{kim2024thermal}. Our algorithm is designed to be generally applicable to all 1D long-range interacting systems with $\alpha>2$. However, when our algorithm is applied to specific Hamiltonians for practical applications, more stringent conditions are typically required, making the reduction of complexity necessary.

Finally, as in the case of short-range interactions~\cite{PRXQuantum.2.040331}, extending the analysis to infinite system setups remains a critical challenge for future research.

\begin{acknowledgments}
All the authors acknowledge the Hakubi projects of RIKEN. 
T. K. was supported by JST PRESTO (Grant No.
JPMJPR2116), ERATO (Grant No. JPMJER2302),
and JSPS Grants-in-Aid for Scientific Research (No.
JP23H01099, JP24H00071), Japan.
Y. K. was supported by the JSPS Grant-in-Aid for Scientific Research (No. JP24K06909). 
This research was conducted during the first author’s internship at RIKEN, which was supervised by the last author.

\end{acknowledgments}

\bibliography{1D_long_simulation}

\begin{thebibliography}{80}
\bibitem [{Foo({\natexlab{a}})}]{Foot0}%
  \BibitemOpen
  \href@noop {} {}We note that when we refer to DMRG in this study, we include
  finite-temperature DMRG and time-dependent DMRG, not solely an algorithm for
  ground-state search.\BibitemShut {Stop}%
\bibitem [{Foo({\natexlab{b}})}]{Foot5}%
  \BibitemOpen
  \href@noop {} {}For example, when applying the cluster expansion
  method~\cite{PhysRevB.91.045138}, we can only obtain the tensor network
  representation for a fully connected graph. However, at low temperatures,
  this approach faces challenges related to the contraction problem.
  Additionally, to apply the HHKL-type decomposition~\cite{PhysRevX.11.011047}
  to long-range interacting systems, the block size needs to be significantly
  larger compared to short-range interactions. For short-range interactions,
  the block size is $\log(1/\epsilon)$, while for long-range interactions, it
  scales as $\poly(1/\epsilon)$. As a result, achieving $\epsilon=1/\poly(n)$
  requires sub-exponential time complexity.\BibitemShut {Stop}%
\bibitem [{Foo({\natexlab{c}})}]{Foot1}%
  \BibitemOpen
\bibfield  {journal} {  }\href@noop {} {}In the general $k$-local case, the
  Hamiltonian $H$ can have at most $n^k$ terms, as it involves interactions
  among at most $k$ sites out of the total $n$ sites. Furthermore, by applying
  the Schmidt decomposition to each interaction term, the maximum bond
  dimension for each term is at most $(d^2)^{\lfloor k/2 \rfloor}$ with
  $\lfloor \cdot \rfloor$ the floor function. Hence, we can construct an MPO
  for each interaction term with a bond dimension less than $d^{k}$. Therefore,
  the total bond dimension required to represent any general $k$-local
  Hamiltonian using an MPO cannot exceed $n^k \times d^{k}$.\BibitemShut
  {Stop}%
\bibitem [{Foo({\natexlab{d}})}]{Foot3}%
  \BibitemOpen
  \href@noop {} {}We typically consider the case of $p=1$ in practice, where
  $\norm{e^{-\beta H}}_{p=1} = Z_\beta$. This norm is important because it
  simplifies to the partition function $Z_\beta$, allowing us to calculate it
  directly from $\tr(M_\beta)$ with the accuracy specified by \begin{align}
  Z_\beta(1 - \epsilon ) \leq \tr(M_\beta) \leq Z_\beta(1 + \epsilon ).
  \end{align} It also implies that the error in the approximation of the Gibbs
  state by the normalized MPO is bounded as follows: $
  \norm{M_\beta/\tr(M_\beta) - \rho_{\beta}}_1 \le 2\epsilon . $ Moreover, for
  any arbitrary observable $O$, the deviation of the expected value computed
  using $M_\beta$ from that computed using the true Gibbs state $\rho_{\beta}$
  is bounded by: $| \tr(O M_\beta)/\tr(M_\beta) - \tr\br{O\rho_{\beta}} | \le
  2\epsilon \norm{O}$\BibitemShut {NoStop}%
\bibitem [{Foo({\natexlab{e}})}]{Foot2}%
  \BibitemOpen
  \href@noop {} {}For $\beta \ge n$, the bond dimension of $e^{\beta \poly
  (\ln(n))}$ is worse than the trivial upper bound of $e^{\orderof{n}}$.
  Therefore, the exact diagonalization achieves the desired bond dimension as
  well as the time complexity in the main theorem.\BibitemShut {Stop}%
\bibitem [{Foo({\natexlab{f}})}]{Foot4}%
  \BibitemOpen
  \href@noop {} {}When considering the polynomial approximation of $e^{-\beta
  H}$, the required polynomial degree is at least $\sqrt{\beta
  \ln(1/\epsilon)}$~\cite{TCS-065}, even when $\norm{H} = \mathcal{O}(1)$.
  Additionally, the bond dimension for generic $k$-local Hamiltonians cannot be
  reduced from $\mathrm{poly}(n)$. Thus, for $\epsilon = 1/\mathrm{poly}(n)$,
  the bond dimension of the MPO must have a quasi-polynomial form, typically
  larger than $e^{\ln^{3/2}(n)}$ for general $k$-local
  Hamiltonians.\BibitemShut {Stop}%
\end{thebibliography}

\clearpage
\newpage

\onecolumngrid

\renewcommand\thefootnote{*\arabic{footnote}}

\addtocounter{section}{0}

\setcounter{equation}{0}
\setcounter{section}{0}

\renewcommand{\theequation}{S.\arabic{equation}}

\renewcommand{\thesection}{S.\Roman{section}}

\begin{center}
{\large \bf Supplementary Material for "Efficient Simulation of 1D Long-Range Interacting Systems at Any Temperature"}\\
\vspace*{0.3cm}
Rakesh Achutha$^{1,2}$, Donghoon Kim$^{1}$, Yusuke Kimura$^{1}$ and Tomotaka Kuwahara$^{1,3,4}$ \\
\vspace*{0.1cm}
$^{1}${\small \it Analytical quantum complexity RIKEN Hakubi Research Team, RIKEN Center for Quantum Computing (RQC), Wako, Saitama 351-0198, Japan}\\
 $^{2}${\small \it Department of Computer Science and Engineering, Indian Institute of Technology (Banaras Hindu University), Varanasi, 221005, India}\\
$^{3}${\small \it PRESTO, Japan Science and Technology (JST), Kawaguchi, Saitama 332-0012, Japan}\\
$^{4}${\small \it RIKEN Cluster for Pioneering Research (CPR), Wako, Saitama 351-0198, Japan} 
\end{center}

\tableofcontents

\section{Several basic statements}
\label{sec2}

\subsection{Operator norm of the boundary interaction for long-range  interacting systems when $\alpha > 2$}

We assume in the main text that the boundary interaction is bounded:
\begin{align}
    \norm{\partial h_A} \leq \Tilde{g}, \qquad \forall A \subset \Lambda,
    \label{Asmpt}
\end{align}
where the boundary interaction $\partial h_{A}$ is defined as the interaction acting on both subsystems $A$ and $B$, derived by subtracting the Hamiltonians $H_A$ and $H_B$, which act exclusively on $A$ and $B$, from the total Hamiltonian,
\begin{align}
    \partial h_A := \sum_{Z: Z \cap A \neq \emptyset, Z \cap B \neq \emptyset} h_Z = H - (H_A + H_B).
    \label{bound-int-def}
\end{align}
We demonstrate that this assumption is inherently satisfied for long-range interactions with a power law decay of $\alpha > 2$:
\begin{lemma} \label{lemma_supp_1}
Considering the power-law decay of interactions in the form 
\begin{align}
    \sum_{Z: Z \ni {i,i'}} \norm{h_Z} \leq \frac{J}{|i-i'|^\alpha} \for i\neq i',
\end{align}
we prove that $\tilde{g}$ in Eq.~\eqref{Asmpt} is $\orderof{1}$ as long as $\alpha>2$.
\end{lemma} 

\noindent
\textit{Proof of Lemma~\ref{lemma_supp_1}.}
Let us consider two adjacent subsystems, $A$ and $B$. 
From the definition of $\partial h_A$, we obtain 
\begin{align}
    \norm{\partial h_A} &\leq \sum_{Z: Z \cap A \neq \emptyset, Z \cap B \neq \emptyset} \norm{h_Z} \leq \sum_{i \in A} \sum_{i' \in B} \sum_{Z: Z \ni i, i'} \norm{h_Z} \leq \sum_{i \in A} \sum_{i' \in B} \frac{J}{|i-i'|^\alpha} \leq \sum_{i=1}^{\infty} \sum_{j=1}^{\infty} \frac{J}{(i+j - 1)^\alpha} = J\zeta (\alpha - 1),
\end{align}
where $\zeta(s)$ is the Riemann zeta function. For $\alpha > 2$, we can choose $\tilde{g} = J\zeta(\alpha - 1) = \mathcal{O}(1)$.
This completes the proof. 	$\square$

\subsection{MPO approximation of the 2-local Hamiltonians} \label{sec:MPO approximation of the 2-local Hamiltonians}

We generally obtain the MPO forms of $k$-local Hamiltonians which have the bond dimensions at most of $n^k d^k$.  
In this section, we show that the bond dimension is improved in the specific cases of the $2$-local Hamiltonians.
\begin{lemma} \label{lemma_supp_2}
For a 2-local Hamiltonian of the form
\begin{align}
	H &= \sum_{1 \leq i < i' \leq n} \frac{1}{|i - i'|^\alpha} \sum_{\xi, \xi' = 1}^{d^2} J_{\xi,\xi'} P_{i, \xi} \otimes P_{i', \xi'},
\end{align}
where $\{P_{i, \xi}\}_{\xi = 1}^{d^2}$ are operator bases on site $i$ and $J_{\xi,\xi'}$ are constants, there exists a Hamiltonian $\tilde{H}$ that approximates $H$ with an error bounded by $\norm{H - \tilde{H}} \leq \epsilon_H$, and admits an MPO representation with bond dimension
\begin{align}
	D_{\tilde{H}} = c_H\ln^2(n/\epsilon_H),
\end{align}
where $c_H$ is an $\orderof{1}$ constant. 
\end{lemma}

{~}\\

\noindent
{\bf Remark.}
Here, the MPO form for the Hamiltonian $\tilde{H}$ is not exactly the same as that of the original Hamiltonian $H$.
We hence need to take into account the error between $H$ and $\tilde{H}$ for the simulation of the quantum Gibbs states (see Sec.~\ref{sec:MPO construction for the 2-local cases}).

{~}\\

\noindent
\textit{Proof of Lemma~\ref{lemma_supp_2}.} 
We approximate $r^{-\alpha}$ as an exponential series in accordance with Ref.~\cite[Theorem~3]{BEYLKIN2010131}, as follows\footnotetext{
We cannot directly employ Ref.~\cite[Theorem~5]{BEYLKIN2010131} since it treats only the regime $r<1$.}:
\begin{align}
	&\abs{r^{-\alpha} - \sum_{s\in\mathbb{Z}} e^{\alpha s x} e^{-e^{s x}r }} \le \epsilon \for r\ge 1,  \notag\\
	&x= \frac{2\pi}{\ln(3)+\alpha \ln[1/\cos(1)] + \ln(1/\epsilon)}. 
\end{align}
We then consider the finite truncation as $s\in \mathbb{Z}_{m}= [-m,m]$. 
We obtain 
\begin{align}
\label{sum_s_mathbb_Z}
&\abs{\sum_{s\in\mathbb{Z}} e^{\alpha s x} e^{-e^{s x}r }- \sum_{s\in\mathbb{Z}_m} e^{\alpha s x} e^{-e^{s x}r } }\le \sum_{s<-m} e^{\alpha s x} +  \sum_{s>m} e^{\alpha s x -e^{s x}}.
\end{align}
We assume that $m$ satisfies
\begin{align} 
\alpha m x -e^{m x} \le -\alpha m x  \quad \longrightarrow  \quad e^{m x} \ge 2\alpha m x \quad &\longrightarrow  \quad  
m x \ge - {\rm ProductLog}\br{-\frac{1}{2\alpha}} \notag \\
& \longrightarrow  \quad  m \ge \frac{2}{x} \ln(2\alpha) ,
\end{align}
where ${\rm ProductLog}(x)$ is equivalent to the Lambert W function $W(x)$, defined by $x=W(x)e^{W(x)}$.  

Under the assumption, we reduce the inequality~\eqref{sum_s_mathbb_Z} to 
\begin{align}
&\abs{\sum_{s\in\mathbb{Z}} e^{\alpha s x} e^{-e^{s x}r }- \sum_{s\in\mathbb{Z}_m} e^{\alpha s x} e^{-e^{s x}r } } 
\le2\sum_{s>m} e^{-\alpha s x} = \frac{2e^{-\alpha x m}}{e^{\alpha x}-1} =  \frac{2}{e^{\alpha x}-1} \br{\frac{\epsilon}{2\alpha}}^{2\alpha}
\end{align}
with 
\begin{align}
m=\left \lceil \frac{2}{x} \ln(2\alpha /\epsilon) \right \rceil .
\end{align}
Under the above choice, we obtain 
\begin{align}
&\abs{r^{-\alpha}  - \sum_{s\in\mathbb{Z}_m} e^{\alpha s x} e^{-e^{s x}r } }\le\epsilon+  \frac{2}{e^{\alpha x}-1} \br{\frac{\epsilon}{2\alpha}}^{2\alpha}\le  \zeta \epsilon ,
\end{align}
where $\zeta$ is an $\orderof{1}$ constant for $\alpha\ge 2$. 

Based on this approximation, we define the Hamiltonian $\tilde{H}$ as:
\begin{align}
    \tilde{H} &=  \sum_{1 \leq i < i' \leq n} \sum_{s\in\mathbb{Z}_m} e^{\alpha s x}   e^{-e^{s x}|i - i'| } \sum_{\xi, \xi' = 1}^{d^2} J_{\xi,\xi'} P_{i, \xi} \otimes P_{i', \xi'}.
\end{align}
We then obtain 
\begin{align}
\norm{H-    \tilde{H}}\le \zeta \epsilon \sum_{1 \leq i < i' \leq n}\sum_{\xi, \xi' = 1}^{d^2} J_{\xi,\xi'} \norm{P_{i, \xi} \otimes P_{i', \xi'}} 
\le \bar{J} n^2\zeta \epsilon ,
\end{align}
with 
\begin{align}
\bar{J} := \sum_{\xi, \xi' = 1}^{d^2} |J_{\xi,\xi'}|  , 
\end{align}
where we set the operator bases such that $\norm{P_{i, \xi}}=1$ for $\forall i$ and $\forall \xi$. 
 
To ensure an approximation error of $\epsilon_H$ for $\norm{H-    \tilde{H}}$, we let $\bar{J} n^2\zeta \epsilon=\epsilon_H$ 
(or $\epsilon=\epsilon_H/(\bar{J} \zeta n^2)$), which implies 
\begin{align}
m= \left \lceil \frac{2}{x} \ln(2\alpha /\epsilon) \right \rceil \le c'_H\ln^2(n/\epsilon_H),
\end{align}
where $c'_{H}$ is a constant of order $\mathcal{O}(1)$.

Moreover, it has been shown that each Hamiltonian of the form $\sum_{1 \leq i < i' \leq n} e^{\alpha s x}  e^{-e^{s x} |i - i'| } \sum_{\xi, \xi' = 1}^{d^2} J_{\xi,\xi'} P_{i, \xi} \otimes P_{i', \xi'}$ 
with exponential decay admits an MPO representation with a bond dimension of $\mathcal{O}(d^2)$~\cite{Pirvu_2010}. Therefore, the MPO representation of $\tilde{H}$ can be constructed by combining the MPOs of each term, resulting in a total bond dimension of:
\begin{align}
	D_{\tilde{H}} &= c_H\ln^2(n/\epsilon_H)
\end{align}
for $d=\orderof{1}$, where $c_H$ is an $\mathcal{O}(1)$ constant.
This completes the proof. $\square$

\section{Proof of main propositions and technical details}

\subsection{Proof of Proposition 1 in the main text}

We now present the proof of Proposition~1 in the main text.
For the convenience of readers, we recall the setup. 
We define a merging operator $\Psi$ that connects two subsets $A$ and $B$ as follows:
\begin{align}
    \Psi = e^{-\beta_0 H_{AB}} e^{\beta_0 (H_A + H_B)}, \qquad H_{AB} = H_A + H_B + \partial h_A,
\end{align}
which leads to the relation
\begin{align}
    e^{-\beta_0 H_{AB}} = \Psi \, e^{-\beta_0 H_A} e^{-\beta_0 H_B}.
\end{align}

\begin{prop}\label{prop_supp_1}
Consider the polynomial expansion of the merging operator $\Psi$ in powers of $\beta_0$, given by:
\begin{align}
    \Psi = e^{-\beta_0 H_{AB}} e^{\beta_0 (H_{A} + H_{B})} 
    &= e^{-\beta_0 H_{AB}} e^{\beta_0 (H_{AB}- \partial h_{A})}  \notag \\
    &= \sum_{m=0}^{\infty} \beta_0^m \sum_{s_1 + s_2 = m} \frac{1}{s_1! s_2!} (-H_{AB})^{s_1} (H_{AB} - \partial h_A)^{s_2}. \label{supp_power}
\end{align}
We truncate the above series to $m_0$ terms and denote the resulting truncated series as $\tilde{\Psi}$:
\begin{align}
    \tilde{\Psi} &= \sum_{m=0}^{m_0} \beta_0^m \sum_{s_1 + s_2 = m} \frac{1}{s_1! s_2!} (-H_{AB})^{s_1} (H_{AB} - \partial h_A)^{s_2}. \label{tilde_psi_def}
\end{align}
Then, if $\beta_0$ is smaller than $\frac{1}{24gk^2}$, we achieve the approximation
\begin{align}
\norm{\Psi - \tilde{\Psi}} \leq \delta_0 \for m_0 \ge \log_2 \br{\frac{c_0}{\delta_0}} , \label{main/prop_supp_1}
\end{align}
where $c_0 =e^{\tilde{g}/(6gk^2)} $ is a constant of $\mathcal{O}(1)$.
\end{prop}

\noindent
\textit{Proof of Proposition~\ref{prop_supp_1}.} 
We derive the approximation of the truncation of the expansion. 
To this end, we first reformulate each term in the expansion~\eqref{supp_power}  by employing an alternative expansion, specifically using the interaction picture:
\begin{align}
    e^{\beta_0 (H_{AB} - \partial h_A)} = e^{\beta_0 H_{AB}} \mathcal{T} \left( e^{-\int_0^{\beta_0} \partial h_x \, dx} \right),
\end{align}
where we define $\partial h_x := e^{- x H_{AB}} \partial h_A e^{x H_{AB}}$ and $\mathcal{T}$ is the time-ordering operator. We then obtain
\begin{align}
    \Psi &= e^{-\beta_0 H_{AB}} e^{\beta_0 (H_{AB}- \partial h_{A})}  \notag \\
    &= \mathcal{T} \left( e^{-\int_0^{\beta_0} \partial h_x \, dx} \right) = \sum_{s = 0}^{\infty} (-1)^{s} \int_{0}^{\beta_{0}} dx_{1} \int_{0}^{x_{1}} dx_{2} \cdots \int_{0}^{x_{s-1}} dx_{s} \, \partial h_{x_{1}} \partial h_{x_{2}} \cdots \partial h_{x_{s}}. \label{supp_interaction_pic}
\end{align}
By applying the decomposition
\begin{align}
	\partial h_x = \sum_{p=0}^{\infty} \frac{(-x)^p}{p!} \mathrm{ad}_{H_{AB}}^p (\partial h_A) =: \sum_{p=0}^{\infty} \frac{(-x)^p}{p!} \partial h^{(p)},
\end{align}
the $m$-th order term in the expansion of $\Psi$ in powers of $\beta_{0}$, which includes the contribution from $\beta_{0}^{m}$ in Eq.~\eqref{supp_interaction_pic}, is expressed as
\begin{align}
	\Psi_{0} &:= 1, \label{supp_interaction_pic_1}\\
	\Psi_m &:= \sum_{s=1}^m \sum_{p_1 + p_2 + \cdots + p_s = m - s} (-1)^s \int_{0}^{\beta_{0}} dx_{1} \int_{0}^{x_{1}} dx_{2} \cdots \int_{0}^{x_{s-1}} dx_{s} \, \frac{(-x_1)^{p_1}}{p_1!} \cdots \frac{(-x_s)^{p_s}}{p_s!} \, \partial h^{(p_1)} \cdots \partial h^{(p_s)} \label{supp_interaction_pic_2}
\end{align}
for $\forall m \geq 1$.
As both Eq.~\eqref{supp_power} and Eq.~\eqref{supp_interaction_pic} are expansions in powers of $\beta_{0}$, the term $\Psi_{m}$ defined above corresponds to the $m$th order term in Eq.~\eqref{supp_power}:
\begin{align}
 \Psi_m =  \beta_0^m \sum_{s_1 + s_2 = m} \frac{1}{s_1! s_2!} (-H_{AB})^{s_1} (H_{AB} - \partial h_A)^{s_2},
\end{align}
which we aim to upper-bound. 

The norm of $\Psi_{m}$ is upper-bounded as
\begin{align}
    \|\Psi_m\| &\leq \sum_{s=1}^m \beta_0^{m-s} \sum_{p_1 + p_2 + \cdots + p_s = m - s} \int_0^{\beta_0} dx_1 \, \int_{0}^{x_{1}} dx_{2} \cdots \int_{0}^{x_{s-1}} dx_{s} \,  \prod_{j=1}^s \frac{1}{p_j!} \|\partial h^{(p_j)}\| \nonumber \\
    &\leq \beta_0^m \sum_{s=1}^m \frac{1}{s!} \sum_{p_1 + p_2 + \cdots + p_s = m - s} \prod_{j=1}^s \frac{1}{p_j!} \|\partial h^{(p_j)}\|. \label{3}
\end{align}
From Theorem 2.1 in \cite{Kuwahara_2016_njp}, if $H$ is a $k$-local and $g$-extensive Hamiltonian and $A$ is a $r$-local operator, then
\begin{align}
    \norm{[H, A]} \leq 6gkr \|A\|. \label{commutationineq}
\end{align}
Since $H_{AB}$ and $\partial h_{A}$ are $k$-local Hamiltonians, $\text{ad}_{H_{AB}}^{p-1} (\partial h_A)$ is at most $kp$-local. Using the inequality \eqref{commutationineq}, we get
\begin{align}
    \|\partial h^{(p)}\| = \|\text{ad}_{H_{AB}}^p (\partial h_A)\| &\leq 6gk(kp) \|\text{ad}_{H_{AB}}^{p-1} (\partial h_A)\| .
\end{align}
By iterating the above process for $\|\text{ad}_{H_{AB}}^{p'} (\partial h_A)\|$ with $p' \leq p-2$ and using \eqref{Asmpt}, we obtain
\begin{align}
    \|\partial h^{(p)}\| \leq (6gk^2)^p p! \tilde{g} =: C^p p! \tilde{g}, \label{supp_norm_hp}
\end{align}
where we define $C := 6gk^2$. Substituting Eq.~\eqref{supp_norm_hp} into Eq.~\eqref{3}, we get
\begin{align}
\label{psim}
    \|\Psi_m\| &\leq \beta_0^m \sum_{s=1}^m \frac{C^{m-s} \tilde{g}^s}{s!} \sum_{p_1 + p_2 + \cdots + p_s = m - s} 1 = \beta_0^m \sum_{s=1}^m \frac{C^{m-s} \tilde{g}^s}{s!} \left(\hspace{-0.35em}\binom{m-s}{s}\hspace{-0.35em}\right).
\end{align}
Using $\left(\hspace{-0.25em}\binom{m-s}{s}\hspace{-0.25em}\right) = \binom{m - 1}{s} \leq 2^{m}$ ,
\begin{align}
	\|\Psi_m\| &\leq (2C\beta_0)^m \sum_{s=1}^{m} \frac{1}{s!} \br{\frac{\tilde{g}}{C}}^{s} \leq (2C\beta_0)^m e^{\tilde{g}/C}. 
\end{align}
Given the assumption $ \beta_0 \leq \frac{1}{24gk^2} = \frac{1}{4C}$, it follows that
\begin{align}
	\|\Psi_m\| \leq 2^{-m} e^{\tilde{g}/C}, \label{psim__2}
\end{align}
which exhibits exponential decay with respect to $m$. 
Thus, the truncation error resulting from approximating $\Psi$ by $\tilde{\Psi}$ is bounded by
\begin{align}
    \norm{\Psi - \tilde{\Psi}} \leq \sum_{m=m_0+1}^\infty \norm{\Psi_m} \leq \sum_{m = m_{0} + 1}^{\infty} 2^{-m} e^{\tilde{g}/C} = e^{\tilde{g}/C} 2^{-m_0}. 
\end{align}
Letting $c_0 =e^{\tilde{g}/C} = e^{\tilde{g} / (6gk^{2})}$, to achieve the desired error $\|\Psi - \tilde{\Psi}\| \leq \delta_{0}$, it is sufficient to choose $m_0$ such that $\|\Psi - \tilde{\Psi}\| \le c_0 2^{-m_0} \leq \delta_0$, i.e.,
\begin{align}
	m_0 \ge \log_2 \br{\frac{c_0}{\delta_0}}.
\end{align}
This completes the proof of Proposition~\ref{prop_supp_1}. $\square$


\subsection{Proof of Proposition 2 in the main text}

In this subsection, we prove Proposition 2 in the main text, which concerns the approximation of the merged quantum Gibbs state.
\\
\begin{prop}\label{prop_supp_2}
Let $M_{2s - 1}^{(q - 1)}$ and $M_{2s}^{(q - 1)}$ represent the MPOs approximating $e^{-\beta_{0} H_{2s - 1}^{(q-1)}}$ and $e^{-\beta_{0} H_{2s}^{(q-1)}}$, respectively, such that
\begin{align}
    \norm{e^{-\beta_0 H_{2s-1}^{(q-1)}} - M_{2s-1}^{(q-1)}}_p &\leq \epsilon_{q-1} \norm{e^{-\beta_0 H_{2s-1}^{(q-1)}}}_p, \\
    \norm{e^{-\beta_0 H_{2s}^{(q-1)}} - M_{2s}^{(q-1)}}_p &\leq \epsilon_{q-1} \norm{e^{-\beta_0 H_{2s}^{(q-1)}}}_p,
\end{align}
for an arbitrary Schatten $p$-norm. 
Using the merging operator $\tilde{\Psi}_{s}^{(q-1)}$, which provides an approximation to $\Psi_{s}^{(q-1)}$ with the error bounded by $\|\Psi_{s}^{(q-1)} - \tilde{\Psi}_{s}^{(q-1)}\| \leq \delta_0$,
the merged quantum Gibbs state $e^{-\beta_{0} H_s^{(q)}}$ can be approximated by the newly constructed MPO $M_s^{(q)}$ based on $M_s^{(q)} = \Tilde{\Psi}_{s}^{(q-1)} M_{2s-1}^{(q-1)} M_{2s}^{(q-1)}$ as
\begin{align}
	\norm{e^{-\beta_{0} H_{s}^{(q)}} - M_{s}^{(q)}}_p \leq \epsilon_{q} \norm{e^{-\beta_{0} H_{s}^{(q)}}}_p, \label{prop_supp_2_main_ineq}
\end{align}
with the error
\begin{align}
	\epsilon_q = a_2 \delta_0 + a_1 \epsilon_{q-1}. \label{prop_supp_2_main_ineq_2}
\end{align}
Here, $a_1$ and $a_2$ are $\mathcal{O}(1)$ constants with $a_{1} > 1$, defined as
\begin{align}
\label{prop_supp_2_a_12}
    a_1 &:= 12 e^{\tilde{g}/(4gk^2)}  , \qquad a_2 := 2 e^{\tilde{g}/(24gk^2)} .
\end{align}
\end{prop}

\noindent
\textit{Proof of Proposition~\ref{prop_supp_2}.} The merging operator is defined as $\Psi_{s}^{(q-1)} = e^{-\beta_0 H_s^{(q)}} e^{\beta_0 (H_{2s - 1}^{(q-1)} + H_{2s}^{(q-1)})}$, and the approximate merging operator $\tilde{\Psi}_{s}^{(q-1)}$ is given by Eq.~\eqref{tilde_psi_def}.
With the MPO defined as $M_s^{(q)} = \tilde{\Psi}_{s}^{(q-1)} M_{2s-1}^{(q-1)} M_{2s}^{(q-1)}$, we proceed with the following inequality:
\begin{align}
    \norm{e^{-\beta_0 H_s^{(q)}} - M_s^{(q)}}_p &= \norm{\Psi_{s}^{(q-1)} e^{-\beta_0 H_{2s-1}^{(q-1)}} e^{-\beta_0 H_{2s}^{(q-1)}} - \tilde{\Psi}_{s}^{(q-1)} M_{2s-1}^{(q-1)} M_{2s}^{(q-1)}}_p \nonumber \\
    &\leq \norm{\br{\Psi_{s}^{(q-1)} - \tilde{\Psi}_{s}^{(q-1)}} e^{-\beta_0 H_{2s-1}^{(q-1)}} e^{-\beta_0 H_{2s}^{(q-1)}} }_p \nonumber \\
    &\qquad + \norm{\tilde{\Psi}_{s}^{(q-1)} \br{e^{-\beta_0 H_{2s-1}^{(q-1)}} e^{-\beta_0 H_{2s}^{(q-1)}}  - M_{2s-1}^{(q-1)} M_{2s}^{(q-1)}}}_p \nonumber \\
    &\leq \norm{\Psi_{s}^{(q-1)} - \tilde{\Psi}_{s}^{(q-1)}}_\infty \norm{e^{-\beta_0 H_{2s-1}^{(q-1)}} e^{-\beta_0 H_{2s}^{(q-1)}} }_p \nonumber \\
    &\qquad + \norm{\tilde{\Psi}_{s}^{(q-1)}}_\infty \norm{e^{-\beta_0 H_{2s-1}^{(q-1)}} e^{-\beta_0 H_{2s}^{(q-1)}} - M_{2s-1}^{(q-1)} M_{2s}^{(q-1)}}_p.\label{starting_ineq_lemma_supp4}
\end{align}
From Ref.~\cite[Lemma 7 therein]{kimura2024clustering}, we can prove for arbitrary operators $\mA$ and $\mB$ 
\begin{align}
\norm{e^{\mA+\mB} -e^{\mA}}_p \le e^{\norm{\mB}} \norm{\mB}  \norm{e^{\mA}}_p . 
\end{align}
By applying the inequality and substituting $-\beta_0 H_s^{(q)}$ for $\mA$ and $\beta_0 H_s^{(q)} - \beta_0 H_{2s}^{(q-1)} -\beta_0 H_{2s-1}^{(q-1)}$ for $\mB$, we obtain $\norm{\mB}\le \tilde{g}$ from assumption \eqref{Asmpt} and
\begin{align}
&\norm{e^{-\beta_0 H_{2s-1}^{(q-1)}} e^{-\beta_0 H_{2s}^{(q-1)}} - e^{-\beta_0 H_s^{(q)}}}_p \leq e^{\beta_0 \tilde{g}} \norm{e^{-\beta_0 H_s^{(q)}}}_p.
\end{align}
The triangle inequality leads to
\begin{align}
\norm{e^{-\beta_0 H_{2s-1}^{(q-1)}} e^{-\beta_0 H_{2s}^{(q-1)}}}_p 
\le \br{e^{\beta_0 \tilde{g}}+1}\norm{e^{-\beta_0 H_s^{(q)}}}_p
\le 2 e^{\beta_0 \tilde{g}} \norm{e^{-\beta_0 H_s^{(q)}}}_p . \label{upp_q-1_to_q}
\end{align}
From the above inequality and the condition $\|\Psi_{s}^{(q-1)} - \tilde{\Psi}_{s}^{(q-1)}\|_p \leq \delta_0$, we reduce inequality~\eqref{starting_ineq_lemma_supp4} to 
\begin{align}
\label{ineq2_lemma_supp4}
    \norm{e^{-\beta_0 H_s^{(q)}} - M_s^{(q)}}_p 
    &\leq  2\delta_0  e^{\beta_0 \tilde{g}} \norm{e^{-\beta_0 H_s^{(q)}}}_p + \norm{\tilde{\Psi}_{s}^{(q-1)}}_\infty \norm{e^{-\beta_0 H_{2s-1}^{(q-1)}} e^{-\beta_0 H_{2s}^{(q-1)}} - M_{2s-1}^{(q-1)} M_{2s}^{(q-1)}}_p.
\end{align}

To estimate the second term on the RHS of \eqref{ineq2_lemma_supp4},
we use the following mathematical lemma:

{~}

\begin{lemma} \label{Claim_O_1O_2}
If two operators $O_1$ and $O_2$ act on disjoint supports, with their respective approximations $\tilde{O}_1$ and $\tilde{O}_2$ defined on the supports of $O_1$ and $O_2$, satisfying 
\begin{align}
    \norm{O_1 - \tilde{O}_1}_p &\leq \epsilon \norm{O_1}_p, \label{Claim_cond1} \\
    \norm{O_2 - \tilde{O}_2}_p &\leq \epsilon \norm{O_2}_p, \label{Claim_cond2}
\end{align}
with $\epsilon<1$, we then obtain 
\begin{align}
    \norm{O_1 O_2 - \tilde{O}_1 \tilde{O}_2}_p \leq 3 \epsilon \norm{O_1 O_2}_p.
\end{align}
\end{lemma}

\noindent
\textit{Proof of Lemma~\ref{Claim_O_1O_2}.} When $O_1$ and $O_2$ operate on disjoint supports, we have $\norm{O_1 O_2}_p = \norm{O_1}_p \norm{O_2}_p$. By leveraging this fact along with Eqs.~\eqref{Claim_cond1} and \eqref{Claim_cond2}, we derive
\begin{align*}
    \norm{O_1 O_2 - \tilde{O}_1 \tilde{O}_2}_p &\leq \norm{O_1 (O_2 - \tilde{O}_2)}_p + \norm{ (O_1 - \tilde{O}_1) \tilde{O}_2}_p \\
    &= \norm{O_1}_p \norm{O_2 - \tilde{O}_2}_p + \norm{O_1 - \tilde{O}_1}_p \norm{\tilde{O}_2}_p \\
    &\leq \epsilon \norm{O_1}_p \norm{O_2}_p + \epsilon \norm{O_1}_p \norm{\tilde{O}_2}_p.
\end{align*}
The relation $\norm{\tilde{O}_{2}}_{p} \leq \norm{O_{2}}_{p} + \norm{O_{2} - \tilde{O}_{2}}_{p} \leq (1 + \epsilon) \norm{O_{2}}_{p}$ and $\epsilon^{2} \leq \epsilon$ leads to
\begin{align*}
    \norm{O_1 O_2 - \tilde{O}_1 \tilde{O}_2}_p &\leq \epsilon \norm{O_1}_p \norm{O_2}_p + \epsilon (1 + \epsilon) \norm{O_1}_p \norm{O_2}_p  = (2 \epsilon + \epsilon^{2}) \norm{O_1}_p \norm{O_2}_p \leq 3 \epsilon \norm{O_1}_p \norm{O_2}_p.
\end{align*}
This completes the proof of Lemma 3. 
	$\square$

{~}\\

Since $e^{-\beta_0 H_{2s-1}^{(q-1)}}$ and $e^{-\beta_0 H_{2s}^{(q-1)}}$ act on disjoint supports (i.e., $L_{2s-1}^{(q-1)} \cap L_{2s}^{(q-1)} = \emptyset$), applying the Lemma 3 yields:
\begin{align}
    \norm{e^{-\beta_0 H_{2s-1}^{(q-1)}} e^{-\beta_0 H_{2s}^{(q-1)}} - M_{2s-1}^{(q-1)} M_{2s}^{(q-1)}}_p &\leq 3 \epsilon_{q-1} \norm{e^{-\beta_0 H_{2s-1}^{(q-1)}} e^{-\beta_0 H_{2s}^{(q-1)}}}_p \notag \\
    &\leq 6 \epsilon_{q-1} e^{\beta_0 \tilde{g}} \norm{e^{-\beta_0 H_{s}^{(q)}}}_p,
\end{align}
where we use the upper bound~\eqref{upp_q-1_to_q} in the second inequality. 

We then upper-bound the norm of  $\tilde{\Psi}_{s}^{(q-1)}$, which is the truncated series up to $m_0$ terms in the expansion of $\Psi_{s}^{(q-1)}$ as in Eq.~\eqref{tilde_psi_def}.
By using the upper bound~\eqref{psim__2}, which gives $ \|\Psi_m\|  \leq 2^{-m} e^{\tilde{g}/C}$, we obtain  
\begin{align}
    \norm{\tilde{\Psi}_{s}^{(q-1)}} \leq \sum_{m = 0}^{m_{0}} \norm{\Psi_{m}} \leq \sum_{m=0}^{m_0} 2^{-m} e^{\tilde{g}/C} \le 2 e^{\tilde{g}/C}=2 e^{\tilde{g}/(6gk^2)} .  
\end{align}
We thus derive the upper bound 
\begin{align}
\norm{\tilde{\Psi}_{s}^{(q-1)}}_\infty \norm{e^{-\beta_0 H_{2s-1}^{(q-1)}} e^{-\beta_0 H_{2s}^{(q-1)}} - M_{2s-1}^{(q-1)} M_{2s}^{(q-1)}}_p
\le  12 e^{\tilde{g}/(6gk^2)+\beta_0 \tilde{g}} \epsilon_{q-1} \norm{e^{-\beta_0 H_{s}^{(q)}}}_p .
\end{align}
By combining the above inequality with~\eqref{ineq2_lemma_supp4}, we have 
\begin{align}
\label{ineq3_lemma_supp4}
    \norm{e^{-\beta_0 H_s^{(q)}} - M_s^{(q)}}_p 
    &\leq  \br{ 2\delta_0  e^{\beta_0 \tilde{g}} + 12 e^{\tilde{g}/(6gk^2)+\beta_0 \tilde{g}} \epsilon_{q-1}  } \norm{e^{-\beta_0 H_s^{(q)}}}_p  \notag \\
    &\leq \br{ 2\delta_0  e^{\tilde{g}/(24gk^2)} + 12 e^{\tilde{g}/(4gk^2)} \epsilon_{q-1}  } \norm{e^{-\beta_0 H_s^{(q)}}}_p   ,
    \end{align}
where we use $\beta_0 \le 1/(24gk^2)$.  
Thus, we establish the desired inequality~\eqref{prop_supp_2_main_ineq}, along with Eqs.~\eqref{prop_supp_2_main_ineq_2} and \eqref{prop_supp_2_a_12}.
This completes the proof of Proposition 2. 
	$\square$

\subsection{Explicit estimation of the bond dimension} \label{sec:Explicit estimation of the bond dimension}

From Proposition 2, we see that $\epsilon_q$ follows a recursive relation, with $\epsilon_1 = 0$ since the blocks in layer 1 have exact MPOs. Using $\epsilon_1=0$, the recursive relation yields the following relations:
\begin{align}
    \epsilon_q &= a_2 \delta_0 + a_1 \epsilon_{q-1} = a_2 \delta_0 \left(1 + a_1 + a_1^2 + \cdots + a_1^{q-2}\right) + a_1^{q-1}\epsilon_1 = a_2 \delta_0 \left(\frac{a_1^{q-1} - 1}{a_1 - 1}\right) + 0  \leq a_2 \delta_0 q\,a_1^{q-2}. \label{supp_recursive_int}
\end{align}
Given that $e^{-\beta_{0} H_{1}^{(q_0)}} = e^{-\beta_{0} H}$ and $M_{\beta_{0}} := M_{1}^{(q_{0})}$, Proposition 2 implies
\begin{align}
	\norm{e^{-\beta_0 H} - M_{\beta_0}}_p &\leq \epsilon_{q_0} \norm{e^{-\beta_0 H}}_p.
\end{align}
Thus, using the inequality~\eqref{supp_recursive_int} and $q_0= \log_2 (n)$, we obtain the MPO approximation of the high-temperature Gibbs state $e^{-\beta_0 H}$ given by:
\begin{align}
    \norm{e^{-\beta_0 H} - M_{\beta_0}}_p &\leq a_2 \delta_0 q_0 a_1^{q_0-2} \norm{e^{-\beta_0 H}}_p
     \leq a_2 \delta_0  n^{\log_2 (2a_1)} \norm{e^{-\beta_0 H}}_p.
    \label{epsilonbeta0}
\end{align}

Starting from the approximate MPO $M_{\beta_{0}}$ for the high-temperature Gibbs state, we construct the MPO for the low-temperature Gibbs state. By concatenating the MPO $M_{\beta_{0}}$ $(\beta / \beta_0)$ times, we obtain $(M_{\beta_0})^{\beta / \beta_0} \approx e^{-\beta H}$.
To derive the error bound for the low-temperature Gibbs state, we begin by considering the approximation in the $(pQ)$-norm, where $Q = \beta / \beta_0$. Given that we have already established the MPO approximation for any general Schatten $p$-norm, it follows that
\begin{align}
    \norm{e^{-\beta_0 H} - M_{\beta_0}}_{pQ} &\leq a_2 \delta_0  n^{\log_2 (2a_1)}  \norm{e^{-\beta_0 H}}_{pQ}.
\end{align}
We now use the inequality from \cite[ineq. (C6) therein]{PhysRevX.11.011047}, which states:
For fixed positive integers \( p_1 \) and \( p_2 \), if \( \norm{e^{-\beta_0 H} - M_{\beta_0}}_{p_1 p_2} \leq \epsilon' \norm{e^{-\beta_0 H}}_{p_1 p_2} \), then 
\begin{equation}
    \norm{e^{-p_1 \beta_0 H} - M_{\beta_0}^{p_1}}_{p_2} \leq \frac{3e}{2} p_1 \epsilon' \norm{e^{-p_1 \beta_0 H}}_{p_2}.
\end{equation}
Applying this to our case, with $p_1 =Q= \beta / \beta_0$ and $p_2 = p$, we get:
\begin{align}
    \norm{e^{-\beta_0 Q H} - (M_{\beta_0})^{Q}}_p &\leq \frac{3e}{2} Q a_2 \delta_0  n^{\log_2(2a_1)}   \norm{e^{-\beta_0 Q H}}_p \leq 5 \left(\frac{\beta}{\beta_0}\right) a_2 \delta_0  n^{\log_2(2a_1)}  \norm{e^{-\beta_0 Q H}}_p.
\end{align}
Now we choose
\begin{align}
    \delta_0 = \frac{\beta_0\epsilon }{5\beta a_2  n^{\log_2 (2a_1)}}\label{delta0}.
\end{align}
Since $a_2$, $\beta_0$, and $a_1$ are $\mathcal{O}(1)$ constants, we have $\delta_0 = \epsilon/\mathrm{poly}(n)$ under the assumption $\beta < n$. By selecting $\delta_0$ as in Eq.~\eqref{delta0}, we obtain the final desired inequality,
\begin{align}
     \norm{e^{-\beta H} - M_{\beta}}_p &\leq \epsilon \norm{e^{-\beta H}}_p.
\end{align}

In the following, we denote the bond dimensions of the MPOs $M_{\beta_0}$ and $M_{\beta}$ by $D_{\beta_0}$ and $D_\beta$, respectively.
The MPO $M_{\beta_0}$ is constructed by $(q_0-1)$ approximate merging processes of $\tilde{\Psi}$ starting from the layer 1.  
The bond dimension of the MPO for the Gibbs states in layer 1 is $d^2$. This bond dimension increases by $\tilde{D}_{\delta_0}$ with the application of the merging operator $\tilde{\Psi}$ at each subsequent layer, where $\tilde{D}_{\delta_0}$ is defined as the bond dimension of $\tilde{\Psi}$ in Eq.~(12).
Hence, the final bond dimension after $(q_0-1)$ mergings becomes $d^2 \tilde{D}_{\delta_0}^{q_0-1}$.
From Eq.~(12), we have $\tilde{D}_{\delta_0} = (m_0 + 1)^2 D_H^{m_0} \leq D_{H}^{2m_{0}}$, where $m_0$ is chosen as 
$$m_0 = \log_2(c_0 / \delta_0)
= \log_2\br{\frac{5 c_0 \beta a_2  n^{\log_2 (2a_1)} }{\beta_0\epsilon } }
$$ according to Proposition 1, with $a_0 > 1$ ensuring that $m_0$ is an integer.
This yields
\begin{align}
    D_{\beta_0} &= d^2 \tilde{D}_{\delta_0}^{q_0-1}  \leq  \tilde{D}_{\delta_0}^{q_0} \leq D_H^{2 m_0 q_0}.
\end{align}
Using Eq.~\eqref{delta0}  and $q_0= \log_2 (n)$, we get
\begin{align}
2 m_0 q_0 &\leq 2\log_2\br{\frac{5 c_0 \beta a_2  n^{\log_2 (2a_1)} }{\beta_0\epsilon } } \log_2 (n) \notag 
    \\
    &= \frac{2}{\ln^2(2)} \ln \br{\frac{5 c_0 \beta a_2  n^{\log_2 (2a_1)} }{\beta_0\epsilon } } \ln (n) \notag 
    \\
    &\leq 5 \brr{\ln \br{\frac{5 c_0a_2}{\beta_0}}+ \log_2 (4a_1)\ln(n/\epsilon)} \ln (n) \notag 
    \\
    &\leq 5 \brr{\ln \br{\frac{5 c_0a_2}{\beta_0}}+ \log_2 (4a_1)} \ln^2 (n/\epsilon)=:b_1 \ln^2 (n/\epsilon) \for (n/\epsilon)\ge e,
\end{align}
with
\begin{align}
b_1:=  5 \brr{\ln \br{\frac{5 c_0a_2}{\beta_0}}+ \log_2 (4a_1)} ,
\end{align}
where we use $\beta < n$ and define an $\mathcal{O}(1)$. 
We recall the definitions of  $\beta_0$, $c_0$, $a_1$ and $a_2$ as follows:
$$
\beta_0 \le \frac{1}{24gk^2},\quad c_0 =e^{\tilde{g}/(6gk^2)} ,\quad a_1:= 12 e^{\tilde{g}/(4gk^2)}  , \quad a_2 := 2 e^{\tilde{g}/(24gk^2)} . 
$$
Hence, the order of $D_{\beta_0}$ becomes $D_H^{\mathcal{O}(\ln^2(n/\epsilon))}$, which directly leads to the following bond dimension $D_\beta$ for the MPO $M_\beta$ corresponding to the low-temperature Gibbs state:
\begin{align}
    D_{\beta}&= D_{\beta_0}^{\beta / \beta_0} \leq D_H^{b_1(\beta/\beta_0 \ln^2(n/\epsilon))} \leq D_H^{(b_1/\beta_0)\beta \ln^2(n/\epsilon)}  \for (n/\epsilon)\ge e.
\end{align}
Hence the order of the bond dimension of $M_\beta$ is $D_H^{\orderof{\beta \ln^2 (n / \epsilon)}}$ since $b_1, \beta_0$ are $\mathcal{O}(1)$ constants. 

\subsection{Estimation of the time complexity}

We finally estimate the time complexity of constructing the MPO $M_\beta$. We begin with the construction of $M_{\beta_0}$. Given two MPOs $M_1$ and $M_2$ with bond dimensions $D_1$ and $D_2$, respectively, the time cost to calculate $M_1 M_2$ is given by $n (D_1 D_2)^2 d^3$.
The total number of merging processes is given by $$2^{q_0-1} + 2^{q_0-2} + \cdots + 1 = 2^{q_0} - 1 < n,$$ and each merging process requires a time cost of at most $n D_\beta^4 d^3$, 
where we use $D_{\beta_0}$ is smaller than $D_{\beta}$, i.e.,  the final bond dimension of the MPO $M_\beta$. 
Thus, the total time cost for constructing $M_{\beta_0}$ is bounded by $n^2 D_\beta^4 d^3$.

For each multiplication of $M_{\beta_0}$ and $M_{\beta_0}^{s}$ with $s \leq \beta/\beta_0$, the time cost is also upper-bounded by $n D_\beta^4 d^3$. Therefore, the time cost to calculate $(M_{\beta_0})^{\beta/\beta_0}$ is at most $n (\beta/\beta_0) D_\beta^4 d^3$.
In total, the time cost to construct $M_\beta$ is upper-bounded by
\begin{align}
    \left(n^2 +  \frac{n\beta}{\beta_0}\right) D_\beta^4 d^3 = D_H^{\orderof{\beta \ln^2 (n/\epsilon)}}.  
\end{align}
By applying $D_H = n^k d^k$ in general cases and $D_H = c_H \ln^2(n/\epsilon)$ in 2-local cases, we confirm the desired time complexity.

\subsection{MPO construction for the 2-local cases} \label{sec:MPO construction for the 2-local cases}
In the 2-local case, the original total Hamiltonian $H$ can be approximated by $\tilde{H}$ as an exponential series, following Sec.~\ref{sec:MPO approximation of the 2-local Hamiltonians}, such that $\norm{H - \tilde{H}} \leq \epsilon_H$ for any arbitrary $\epsilon_H$. The MPO for the approximate Hamiltonian $\tilde{H}$ is then constructed according to Sec~\ref{sec:Explicit estimation of the bond dimension}, satisfying
\begin{align}
	\norm{e^{-\beta \tilde{H}} - \tilde{M}_{\beta}}_p \leq \frac{\epsilon}{3} \norm{e^{-\beta \tilde{H}}}_p,\label{2-l,initmpoerror}
\end{align}
where we assume $\epsilon\le1$. 
Here the bond dimension for $\tilde{M}_{\beta}$ is given by $D_H^{(b_1/\beta_0)\beta \ln^2(3n/\epsilon)}$ with $D_H=c_H \ln^2(n/\epsilon_H)$.

We estimate the error between $e^{-\beta H}$ and $e^{-\beta \Tilde{H}}$ using the Ref.~\cite[Lemma 7 therein]{kimura2024clustering}, which states that we can prove for arbitrary operators $\mA$ and $\mB$ 
\begin{align}
\norm{e^{\mA+\mB} -e^{\mA}}_p \le e^{\norm{\mB}} \norm{\mB} \cdot \norm{e^{\mA}}_p . 
\end{align}
By applying the inequality and substituting $-\beta H$ for $\mA$ and $\beta \Tilde{H} - \beta H$ for $\mB$, we get
\begin{align}
    \norm{e^{-\beta H} - e^{-\beta \Tilde{H}}}_p &\leq e^{\beta \norm{H-\Tilde{H}}}\beta \norm{H-\Tilde{H}}\cdot \norm{e^{-\beta H}}_p \leq e^{\beta \epsilon_H}\beta \epsilon_H \norm{e^{-\beta H}}_p \leq 3\beta \epsilon_H\norm{e^{-\beta H}}_p,
\end{align}
where we choose $\epsilon_H$ such that $\epsilon_H\le 1/\beta$. 
For a given $\beta$, if we choose $\epsilon_H \leq \epsilon/ (6\beta)$\footnote{From this inequality, the condition~$\beta \epsilon_H\le 1$ is satisfied because $\epsilon\le 1$.}, we obtain
\begin{align}
    \norm{e^{-\beta H} - e^{-\beta \Tilde{H}}}_p &\leq \frac{\epsilon}{2} \norm{e^{-\beta H}}_p.
    \label{2-l,Hamerror}
\end{align}
Using inequalities~\eqref{2-l,initmpoerror} and~\eqref{2-l,Hamerror},
\begin{align}
     \norm{e^{-\beta \Tilde{H}}- M_{\beta}}_p \leq \frac{\epsilon}{3} \norm{e^{-\beta \tilde{H}}}_p \leq \frac{\epsilon}{3}  \br{\norm{e^{-\beta \Tilde{H}}-e^{-\beta H}}_p + \norm{e^{-\beta H}}_{p}} 
     &\leq \frac{\epsilon}{3} \br{\frac{\epsilon}{2} + 1} \norm{e^{-\beta H}}_{p} \notag \\
     &\le \frac{\epsilon}{2}  \norm{e^{-\beta H}}_{p} ,
     \label{2-l,mpoerror}
\end{align}
where we use $\epsilon\le 1$ to derive $\epsilon/2 + 1\le 3/2$. 

Therefore, using inequalities~\eqref{2-l,Hamerror} and \eqref{2-l,mpoerror}, we finally obtain the following error
\begin{align}
    \norm{e^{-\beta H}- M_{\beta}}_p \leq \norm{e^{-\beta H}-e^{-\beta \Tilde{H}}}_p + \norm{e^{-\beta \Tilde{H}}-M_{\beta}}_p \leq \epsilon \norm{e^{-\beta H}}_p.
\end{align}
This gives us the approximation MPO for the final Gibbs state $e^{-\beta H}$ for the 2-local case.

%
%

\section{Quantum preparation of the quantum Gibbs state}

Finally, we discuss the preparation of quantum Gibbs states on a quantum computer. 
Here, we start with the initial product state and consider how many elementary quantum gates are sufficient to implement the quantum Gibbs states up to an error $\epsilon$.
The problem is called quantum Gibbs sampling, and there are various kinds of quantum algorithms~\cite{PhysRevLett.103.220502,PhysRevLett.105.170405,temme2011quantum,Yung754,PhysRevLett.116.080503,Kastoryano2016,Brandao2019,chen2023}.  
In one-dimensional cases, the time complexity of the quantum Gibbs sampling is derived from the MPO approximation. 
In conclusion, the time complexity (or circuit depth) for this preparation is also given by the quasi-polynomial form, i.e., $e^{C_0 \beta \ln^3(n/\epsilon)}$.

To achieve this, we first purify the Gibbs state into the form of a thermofield double state~\cite{zhu2020generation}, which can also be well-approximated by a Matrix Product State (MPS) with the same bond dimension as that of $M_\beta$~\cite[Sec. VI A therein]{PhysRevX.11.011047}. 
For the quantum circuit representation of the MPS, we refer to Ref.~\cite[Appendix B therein]{PRXQuantum.2.010342}, where it is generally shown that an arbitrary MPS with bond dimension $\chi$ can be constructed using a circuit with depth $n \times \mathrm{poly}(\chi)$. By setting $\chi = e^{C_0 \beta \ln^3(n/\epsilon)}$ for the quantum Gibbs state's preparation, we achieve the desired circuit depth.

\end{document}